\documentclass[12pt]{article}
\usepackage{graphicx}
\usepackage[letterpaper,margin=1in]{geometry}
\usepackage{authblk}
\usepackage{upgreek}
\usepackage{amsmath}
\usepackage{multirow,tabularx}
\usepackage{amssymb}
\usepackage{tablefootnote}
\usepackage{xcolor}
\usepackage{verbatim}
\usepackage{hyperref}
\usepackage{setspace}
\usepackage{nomencl}
\usepackage{etoolbox}
\usepackage[
	backend=biber,
	style=numeric,
	sorting=none
]{biblatex}
\usepackage{lineno}

\makenomenclature

\usepackage{etoolbox}
\renewcommand\nomgroup[1]{%
  \item[\bfseries
  \ifstrequal{#1}{M}{Main text}{%
  \ifstrequal{#1}{P}{Appendix}{}}%
]}

\title{Discrete element method model of soot aggregates}
\author[1]{Egor V. Demidov}
\author[2]{Gennady Y. Gor}
\author[1,2]{Alexei F. Khalizov\footnote{Corresponding author, khalizov@njit.edu}}
\affil[1]{Department of Chemistry and Environmental Science, New Jersey Institute of Technology, 161 Warren Street, Newark, NJ 07103, United States of America}
\affil[2]{Department of Chemical and Materials Engineering, New Jersey Institute of Technology, 161 Warren Street, Newark, NJ 07103, United States of America}

\date{}

\begin{document}

\maketitle

%

\section*{Abstract}

Soot is a component of atmospheric aerosols that affects climate by scattering and absorbing the sunlight. Soot particles are fractal aggregates composed of elemental carbon. In the atmosphere, the aggregates acquire coatings by condensation and coagulation, resulting in significant compaction of the aggregates that changes the direct climate forcing of soot. Currently, no models exist to rigorously describe the process of soot restructuring, reducing prediction accuracy of atmospheric aerosol models. We develop a discrete element method contact model to simulate restructuring of fractal soot aggregates, represented as assemblies of spheres joined by cohesion and by sintered necks. The model is parametrized based on atomic force spectroscopy data and is used to simulate soot restructuring, showing that the fraction of necks in aggregates determines the restructuring pathway. Aggregates with fewer necks undergo local compaction, while aggregates with nearly-full necking prefer global compaction. Additionally, full compaction occurs within tens of nanoseconds, orders of magnitude faster than the timescale of soot aging through condensation. An important implication is that in atmospheric soot aggregates, the rate of condensation determines how many necks are fractured simultaneously, affecting the restructuring pathway, \textit{e.g.}, producing highly compact, thinly-coated soot as observed in recent studies.

\newpage

\printnomenclature

\section{Introduction}

Atmospheric aerosols affect climate directly by scattering and absorbing light \cite{chylek1995effect} and indirectly by enhancing cloud formation \cite{lohmann2005global}. One of the components of atmospheric aerosols and a strong light absorber is soot \cite{haywood2000estimates}. Soot is formed upon combustion of carbon-containing fuels in a multi-step process involving thermolysis of the fuel, PAH formation, nucleation, and oxidation \cite{thomson2018radical,johansson2018resonance}. As a result, fractal-like structures consisting mostly of elemental carbon are produced \cite{demirdjian2007heterogeneities}. The micro-structure of carbonaceous phase can be either amorphous, fullerenic, or a combination of both, depending on conditions under which the aggregate was formed \cite{grieco2000fullerenic}.

While in the atmosphere, soot aggregates are subject to aging through a number of pathways, including vapor condensation to form coatings around aggregates \cite{saathoff2003coating}. In a process akin to adsorption-induced deformation \cite{gor2017adsorption,gor2024perspective}, capillary forces induced by the liquid coatings cause initially fractal aggregates to become more compact \cite{kutz1992characterization,chen2018single}. This process of restructuring changes the radiative forcing properties of the aggregates \cite{sorensen2001light,demidov2024differences,khalizov2009enhanced}. In order to improve the representation of soot in aerosol models, it is necessary to develop a model that can predict the morphology of soot aggregates based on their surrounding conditions and time spent in the atmosphere. The implications of mechanical properties of soot aggregates extend far beyond the study of soot restructuring for environmental considerations. Carbon black (CB), a material similar in nature to soot \cite{long2013carbon}, is widely used as an additive in manufacturing of rubbers, polyurethane coatings, oxide cathodes, inks, etc. Understanding of mechanical properties of carbon black is important for its industrial applications \cite{smallwood1944limiting,medalia1987effect}.

The only attempt to model the mechanics of soot has been undertaken by Schnitzler \textit{et al.} \cite{schnitzler2017coating}, who studied the effect of surface tension of the coating material on the extent of aggregate restructuring. Particle bonding is not explicitly considered in that study; adhesive and liquid-originating interactions are modeled with simple, pairwise force fields. Additionally, Schnitzler's model does not consider rotation of particles, hence it cannot differentiate necked and point-touch contacts that are present in a real soot aggregate (Figure \ref{fig:contact-types}). While Schnitzler's model qualitatively supports the discoveries made in their experimental study, it lacks an accurate representation of necks that are usually present in fresh soot and resist the restructuring process. Another noteworthy study is the discrete element method (DEM) model for humidified TiO\textsubscript{2} nanoparticle aggregate films under mechanical load by Laube \textit{et al.} \cite{laube2018new}, who consider the impact of capillary and solvation forces that occur in pairs of wetted particles. Capillary and solvation forces are modeled using force fields parametrized by the authors based on molecular dynamics simulations. The model also considers friction and rotational degrees of freedom of individual particles. Despite the rigor of Laube's model, their methodology is not directly transferable to modeling soot restructuring because they assumed rigid aggregates where displacement of primary particles in an aggregate relative to each other is prohibited, which is a valid assumption for agglomerates of aggregates in their study, but does not allow for any restructuring within aggregates by definition, and cannot be used in a model for soot where all contacts are necked. Kelesidis \textit{et al.}  \cite{kelesidis2023process} used DEM to model formation of soot by agglomeration of fine unit clusters into larger spherules, which in turn agglomerate into fractal aggregates of primary spherules. Kelesidis \textit{et al.} did not consider the mechanical behavior of aggregates.

Our goal is to develop a multi-scale framework for simulation of soot aging and restructuring. Previously, our group introduced a technique to account for the impact of non-spherical geometry of soot on vapor condensation kinetics \cite{ivanova2020kinetic,chen2018single}. Now, we present a DEM contact model for soot aggregate mechanics. The contact model allows for creation of elastic bonds (necks) and non-bonded contacts within a soot aggregate. Our goal was to create a model robust enough to reproduce an experiment yet computationally efficient to be used to make predictions by integrating it in larger aerosol models. The developed model is used in this work to simulate an atomic force microscopy (AFM) spectroscopy experiment \cite{rong2004complementary}, as well as aggregate restructuring by capillary forces (with a simplistic representation of coating materials). The AFM force spectroscopy simulation is used to parametrize and validate the model. Several restructuring simulations are conducted to investigate the timescale of soot restructuring, the impact of inter-particle bonds on the outcome of restructuring, and to find an optimal parameter to track the progress of a restructuring simulation.

\section{Developed framework and methodology}

\subsection{Numerical model}

\begin{figure}[htp]
    \centering
    \includegraphics[width=0.6\textwidth]{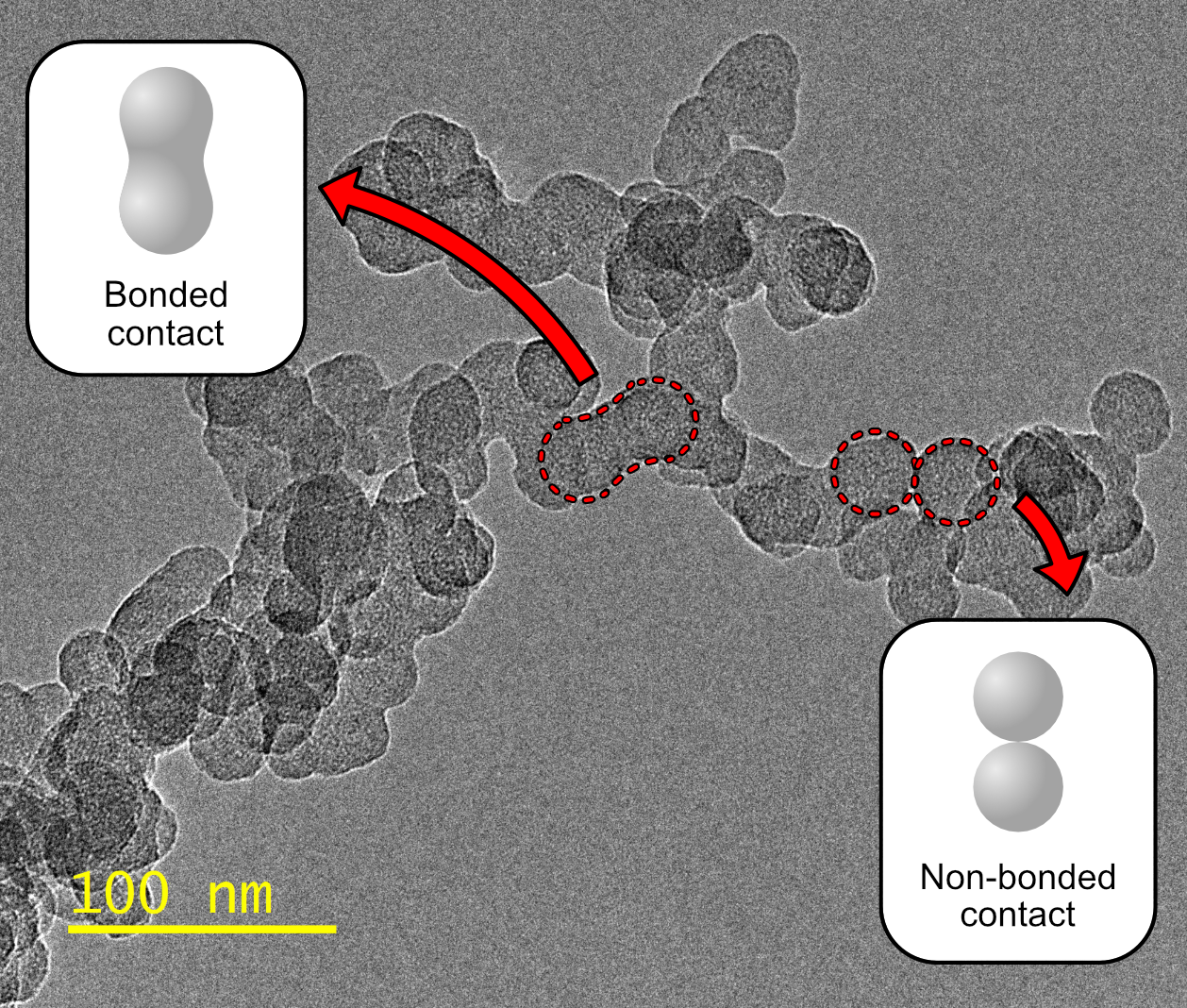}
    \caption{The two possible inter-particle contact types shown on a TEM image of a soot aggregate. Particles that are bonded contain a solid neck between the particles. Particles that are not bonded are touching at one point.}
    \label{fig:contact-types}
\end{figure}

Geometrically, soot aggregates can be approximated as assemblies of spherical particles \cite{gwaze2006comparison}. On the scale of this model, we disregard the micro-structure of primary particles and assume that they are solid spheres. Hence, discrete element method (DEM) is an appropriate simulation technique. DEM was introduced by Cundall and Strack \cite{cundall1979discrete}. It is a numerical technique for simulating mechanical behavior of granular systems. This method is based on computing the resultant force, $\mathbf{f}$\nomenclature[M,f]{\(\mathbf{f}\)}{resultant force}, and torque, $\boldsymbol{\uptau}$\nomenclature[M,t]{\(\boldsymbol{\uptau}\)}{resultant torque}, acting on each particle comprising an aggregate and integrating Newton's second law with respect to time to obtain the time-evolution of the system of particles,
\begin{equation}
    m\ddot{\mathbf{x}}=\mathbf{f}
\end{equation}
\begin{equation}
    I\ddot{\boldsymbol{\uptheta}}=\boldsymbol{\uptau}
\end{equation}
where $m$\nomenclature[M,M,m]{\(m\)}{mass} is mass, $I$\nomenclature[M,M,I]{\(I\)}{moment of inertia} is moment of inertia, $\bf x$\nomenclature[M,M,x]{\(\mathbf{x}\)}{position} is position, and $\boldsymbol{\uptheta}$\nomenclature[M,M,t]{\(\boldsymbol{\uptheta}\)}{orientation} is orientation of the particle. The key assumption here that makes DEM different from other discrete techniques, such as molecular dynamics, is that individual particles are not point masses and can undergo meaningful rotation. Moment of inertia is $I=2/5mr^2$ \nomenclature[M,M,r]{\(r\)}{primary particle radius} for solid spherical primary particles and $I=2/3mr^2$ for hollow spherical primary particles (descriptive of internally oxidized soot  \cite{kelesidis2024oxidation}). Hence, torques and rotational degrees of freedom need to be considered. The source of torques are tractions applied to inter-particle contacts. The tractions are either a result of deformation of an elastic inter-particle bond (in the case of a particle pair joined with a neck, see Figure \ref{fig:contact-types}) or friction and cohesion (in the case of a particle pair engaged in a point-touch contact, see Figure \ref{fig:contact-types}). By using different contact force models, we can simulate aggregates with both necked and point-touch contacts and vary the fraction of necks to see the impact of necks on the outcome of restructuring. The model for force applied to particles in contact determines whether the contacting pair behaves as a bonded union or as non-bonded grains.
A non-bonded contact is a contact that does not have any tensile strength. Such a contact only results in a force when two particles collide with each other. A bonded contact is a contact that does have tensile strength. A pair of particles in which a bonded contact has been established behaves approximately as a rigid body. A small motion of one particle in the pair relative to the other is resisted by the bonding forces. The resultant forces seek to restore the un-deformed configuration of the pair.
In addition to contact forces, there are also body forces that can occur between particles that are not in direct contact with each other, \textit{i.e.} gravity or van der Waals attraction. A body force is applied to the center of mass of a particle and does not directly cause rotation. Gravity is not included in this model because it is irrelevant at the scale of these simulations and cannot have an impact on the restructuring process, because particles in an aggregate accelerated by gravity retain constant velocity relative to each other. In the next section, we develop the general concepts of the contact model that are used in both the rigid bond model and the frictional contact model.

\subsubsection{Development of the contact model}

\label{sec:general-contact-model}

\begin{figure}[htp]
    \centering
    \includegraphics[width=\textwidth]{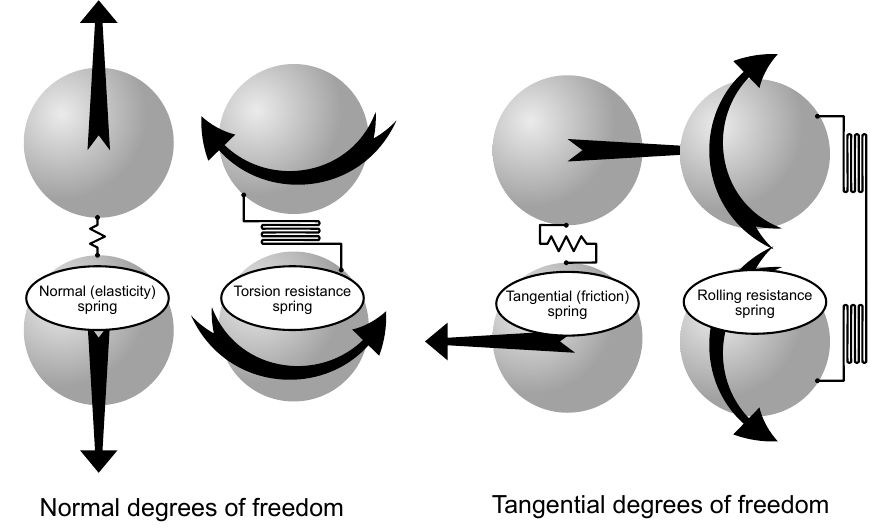}
    \caption{Degrees of freedom of two particles in a common reference frame and illustrations of springs that are used to constrain them}
    \label{fig:dofs}
\end{figure}

In a system of $N$\nomenclature[M,M,N]{\(N\)}{number of primary particles} particles we can have up to $\frac{1}{2}N(N-1)$ binary interactions. Let us pick two particles $i$ and $j$ from system and consider their common reference frame. There are four types of relative motion that particles $i$ and $j$ can undergo, corresponding to the four degrees of freedom that we would like to control (Figure \ref{fig:dofs}). These degrees of freedom are (a) normal translation, (b) torsion, (c) tangential translation, and (d) rolling.
In order to define these degrees of freedom, consider a unit normal vector $\mathbf{n}$\nomenclature[M,n]{\(\mathbf{n}\)}{unit normal vector} that is pointing from the center of particle $i$, $\mathbf{x}_i$, towards the center of particle $j$, $\mathbf{x}_j$:
\begin{equation}
    \mathbf{n}=\frac{\mathbf{x}_j-\mathbf{x}_i}{\lVert \mathbf{x}_j-\mathbf{x}_i\rVert}
\end{equation}
Then the relative velocity at the contact point, $\mathbf{v}_{ij}$\nomenclature[M,vij]{\(\mathbf{v}_{ij}\)}{relative velocity at a contact point}, is:
\begin{equation}
    \mathbf{v}_{ij}=\mathbf{v}_j-\mathbf{v}_i+\boldsymbol{\upomega}_j\times a\mathbf{n}+\boldsymbol{\upomega}_i\times a\mathbf{n}
\end{equation}
where $\mathbf{v}_i$\nomenclature[M,v]{\(\mathbf{v}\)}{translational velocity of a particle} and $\mathbf{v}_j$ are translational velocities of the particles, $\boldsymbol{\upomega}_i$\nomenclature[M,w]{\(\boldsymbol{\upomega}\)}{angular velocity of a particle} and $\boldsymbol{\upomega}_j$ are angular velocities of the particles, $r$ is particle radius (all particles are assumed to have equal radii in this study), and $a$\nomenclature[M,a]{\(a\)}{primary particle radius corrected for overlap/separation} is particle radius corrected for overlap/separation $\delta$\nomenclature[M,d]{\(\delta\)}{overlap/separation between particles} between particles $i$ and $j$:
\begin{equation}
    a=r+\frac{1}{2}\delta
\end{equation}
\begin{equation}
    \delta=\lVert\mathbf{x}_j-\mathbf{x}_i\rVert-2r
\end{equation}
Then $\mathbf{v}_{ij}$ can be decomposed into normal and residual (tangential) components\nomenclature[M,vijn]{\(\mathbf{v}_{ij,\rm n}\)}{normal relative velocity at a contact point}\nomenclature[M,vijt]{\(\mathbf{v}_{ij,\rm t}\)}{tangential relative velocity at a contact point}:
\begin{equation}
    \mathbf{v}_{ij,\rm n}=\left(\mathbf{v}_{ij}\cdot\mathbf{n}\right)\mathbf{n}
\end{equation}
\begin{equation}
    \mathbf{v}_{ij,\rm t}=\mathbf{v}_{ij}-\left(\mathbf{v}_{ij}\cdot\mathbf{n}\right)\mathbf{n}
\end{equation}
Similarly, the relative angular velocity\nomenclature[M,wij]{\(\boldsymbol{\upomega}_{ij}\)}{relative angular velocity at a contact point} of the particles in the pair,
\begin{equation}
    \boldsymbol{\upomega}_{ij}=\boldsymbol{\upomega}_j-\boldsymbol{\upomega}_i
\end{equation}
can be decomposed into normal (torsional)\nomenclature[M,wijn]{\(\boldsymbol{\upomega}_{ij,\rm n}\)}{normal relative angular velocity at a contact point} and residual (rolling)\nomenclature[M,wijr]{\(\boldsymbol{\upomega}_{ij,\rm r}\)}{tangential relative angular velocity at a contact point} components:
\begin{equation}
    \boldsymbol{\upomega}_{ij,\rm n}=\left(\boldsymbol{\upomega}_{ij}\cdot\mathbf{n}\right)\mathbf{n}
\end{equation}
\begin{equation}
    \boldsymbol{\upomega}_{ij,\rm r}=\boldsymbol{\upomega}_{ij}-\left(\boldsymbol{\upomega}_{ij}\cdot\mathbf{n}\right)\mathbf{n}
\end{equation}
Different subscripts are used for $\mathbf{v}_{ij,\rm t}$ and $\boldsymbol{\upomega}_{ij,\rm r}$ to emphasize the fact that their directions are not necessarily the same.

In order to control each degree of freedom listed in Figure \ref{fig:dofs}, we insert four springs between particles $i$ and $j$. The rates of stretching/contraction of these four (normal, torsional, tangential, and rolling) springs are given by $\mathbf{v}_{ij,\rm n}$, $r\boldsymbol{\upomega}_{ij,\rm n}$, $\mathbf{v}_{ij,\rm t}$, and $\boldsymbol{\upomega}_{ij,\rm r}\times a\mathbf{n}$ respectively. Length of the normal spring can be computed directly at every time step from positions of particles $i$ and $j$. As to the remaining three springs, their lengths start at zero when the contact is established and need to be incremented throughout the duration of the contact. Let us define a generic term for a vector representing one of the three springs (torsional, tangential, or rolling), $\boldsymbol\upxi$\nomenclature[M,x]{\(\boldsymbol\upxi\)}{spring vector}, and its rate of change, $\dot{\boldsymbol\upxi}$. Then the algorithm at every time step is as follows:
\begin{enumerate}
    \item Load the spring $\boldsymbol\upxi'$ stored in memory since the previous time step
    \item Compute $\boldsymbol\upxi$ by rotating $\boldsymbol\upxi'$ to be orthogonal to $\mathbf{n}$ and re-scaling the rotated spring to match the magnitude of $\boldsymbol\upxi'$
    \item Increment $\boldsymbol\upxi$ by $\dot{\boldsymbol\upxi}\Delta t$\nomenclature[M,t]{\(\Delta t\)}{integration time step}
    \item Use the value of $\boldsymbol\upxi$ to compute the forces and torques acting on particles $i$ and $j$
    \item Store $\boldsymbol\upxi$ in the variable $\boldsymbol\upxi'$ to be used at the next iteration
\end{enumerate}

\subsubsection{Elastic bond model}

For a pair of particles that is connected with an elastic bond (\textit{i.e.}, two monomers connected by a neck), we would like to approximate a rigid--body motion. In other words, the common reference frame of particles $i$ and $j$ can rotate and translate, but any translation or rotation of particle $i$ relative to particle $j$ should be restricted. The distance between all points in the pair of particles should be approximately preserved over time. It can be shown that when using the springs defined in the previous subsection to restrict the motion of particles $i$ and $j$, then the union of particles $i$ and $j$ will undergo rigid body motion as the stiffness of inserted springs approaches infinity. In the simulation we need to use a finite stiffness, but as long as the amplitude of oscillations is much smaller than the length scale of particles in the simulation, the motion will, approximately, be rigid.

To stabilize the system over time and dissipate any vibrational kinetic energy in the bonds, each spring is supplemented by a dashpot element. The force, $\mathbf{f}$, exerted on particle $i$ by each spring in the $ij$ bond is given by:
\begin{equation}
    \mathbf{f}=k\boldsymbol{\upxi}+\eta\dot{\boldsymbol{\upxi}}
    \label{eq:generic-shear-force}
\end{equation}
where $k$\nomenclature[M,k]{\(k\)}{spring stiffness} is stiffness and $\eta$\nomenclature[M,n]{\(\eta\)}{spring damping coefficient} is the damping coefficient of the respective spring. Forces exerted on particle $j$ are equal in magnitude and opposite in direction. Forces arising from the normal and tangential springs are applied to the particles. The force associated with the tangential spring will also give rise to torques because the tangential force is not collinear with $\mathbf{n}$. Forces computed from the torsion and rolling resistance springs are quasi-forces that are not applied to particles $i$ and $j$, but are only used to compute torques that will be applied to the particles.

Soot restructuring is limited by the strength of elastic bonds between primary particles \cite{weber1997situ}. It is known that most-to-all neighboring primary particles in a soot aggregate are initially bonded with rigid necks \cite{rothenbacher2008fragmentation}, but some of the bonds break under the pressure applied by the AFM tip or the stress imposed by the surface tension of a condensed liquid. Neck fracture parameterization should consider microstructure and possible anisotropy of carbonaceous materials, but that is beyond the scope of this study, which focuses primarily on contact model development. \textit{In lieu} of a bond strength parametrization, we prescribe a certain fraction of broken bonds at the start of every simulation. Then, aggregate fragments can fold, slide, and roll relative to each other only at the non-bonded contacts. If the fraction of broken bonds is zero, then all adjacent primary particles in the aggregate are bonded with necks and no folding, sliding, or rolling is possible.

\subsubsection{Frictional contact model}

The model described by Luding \cite{luding2008cohesive} is used to simulate frictional contacts between non-bonded particles. A brief description is provided here and the reader is referred to Luding \cite{luding2008cohesive} for more details. Luding's model uses the same four springs described in Section \ref{sec:general-contact-model} to compute normal and tangential forces, rolling and torsion resistance torques. Instead of directly setting force proportional to spring elongation, a certain degree of slip is allowed between particles in contact. That is done by computing a test force, $\mathbf{f}_{0}$\nomenclature[M,f0]{\(\mathbf{f}_{0}\)}{frictional test force}, from current spring vector, $\boldsymbol{\upxi}$, and rate of spring stretching, $\dot{\boldsymbol{\upxi}}$, using Equation \ref{eq:generic-shear-force}. Then, based on the magnitude of the test force, a decision is made whether static or dynamic friction should be used,
\begin{equation}
    \text{if}\ \lVert\mathbf{f}_0\rVert\leq f_{\rm C,s}\ \text{use static friction}
\end{equation}
\begin{equation}
    \text{if}\ \lVert\mathbf{f}_0\rVert> f_{\rm C,s}\ \text{use dynamic friction}
\end{equation}
where $f_{\rm C,s}$\nomenclature[M,fcs]{\(f_{\rm C,s}\)}{static friction force} is static friction force. Static and dynamic \nomenclature[M,fcd]{\(f_{\rm C,d}\)}{dynamic friction force} friction forces are computed using Coulomb's law of friction,
\begin{equation}
    f_{\rm C,s}=\mu_{\rm s} f_{\rm n}
\end{equation}
\begin{equation}
    f_{\rm C,d}=\mu_{\rm d} f_{\rm n}
\end{equation}
where $\mu_{\rm s}$\nomenclature[M,ms]{\(\mu_{\rm s}\)}{static friction coefficient} is the static friction coefficient, $\mu_{\rm d}$\nomenclature[M,md]{\(\mu_{\rm d}\)}{dynamic friction coefficient} is the dynamic friction coefficient, and $f_{\rm n}$\nomenclature[M,fn]{\(f_{\rm n}\)}{normal force magnitude at a contact point} is the magnitude of the normal force between the two particles.

In case the contact is determined to be in the state of static friction, the tangential spring $\boldsymbol\upxi$ is incremented as described in Subsection \ref{sec:general-contact-model} and the test force computed using Equation \ref{eq:generic-shear-force} is used to compute torques and, in the case of the tangential force, is also applied to the contacting particles. In case the contact is determined to be in the state of dynamic friction, the spring is not allowed to stretch anymore. Instead, a certain extent of slipping is allowed. The spring to be used at the next iteration, $\boldsymbol\upxi'$, is set to
\begin{equation}
    \boldsymbol\upxi'=-\frac{1}{k}\left(f_{\rm C,d}\frac{\mathbf{f}_0}{\lVert \mathbf{f}_0\rVert}+\eta\dot{\boldsymbol\upxi}\right)
\end{equation}
and the magnitude of the Coulomb's force, $f_{\rm C,d}$, is used to compute torques and, if applicable, accelerate the particles in contact.

The model is only enabled when the normal force is repulsive. Once particles are not overlapping, the frictional force is set to zero and accumulated springs are reset.

\subsubsection{Non-contact forces}

\label{sec:non-contact-forces}

Primary particle diameter of soot from different sources ranges between 10 to 50 nm \cite{adachi2007fractal,wentzel2003transmission}. Primary particles that constitute soot aggregates in this study are assumed to be 28 nm in diameter, which corresponds to the size of primary particles particles in soot used by Enekwizu \textit{et al.} \cite{enekwizu2021vapor} in an experimental study of soot restructuring and by Chen \textit{et al.}  \cite{chen2018single} and Ivanova \textit{et al.} \cite{ivanova2020kinetic} in the study of vapor condensation on soot. At that scale, van der Waals attraction is significant between two particles that are in proximity. Hamaker \cite{hamaker1937london} derived the potential energy, $U$\nomenclature[M,U]{\(U\)}{van der Waals attraction potential energy}, due to van der Waals attraction between two particles of radii $r_1$ and $r_2$ whose surfaces are separated by distance $\delta$ from each other to be:
\begin{equation}
    U=-\frac{A}{6}\left[\frac{2r_1r_2}{(2r_1+2r_2+\delta)\delta}+\frac{2r_1r_2}{(2r_1+\delta)(2r_2+\delta)}+\ln\frac{(2r_1+2r_2+\delta)\delta}{(2r_1+\delta)(2r_2+\delta)}\right]
    \label{eq:hamaker}
\end{equation}
Equation \ref{eq:hamaker} simplifies when all particles are of equal radius $r$:
\begin{equation}
	U=-\frac{A}{6}\left[\frac{2r^2}{(4r+\delta)\delta}+\frac{2r^2}{(2r+\delta)^2}+\ln\frac{(4r+\delta)\delta}{(2r+\delta)^2}\right]
    \label{eq:vdwpe}
\end{equation}
where $A$\nomenclature[M,A]{\(A\)}{Hamaker constant} is the Hamaker constant -- a material property. The magnitude of force acting on the particles can be derived by differentiating potential energy, $U$, with respect to separation distance, $\delta$, and the direction of the force will coincide with the normal unit vector $\bf n$ defined earlier:
\begin{equation}
    \mathbf{f}=-\frac{A}{6}\left[\frac{(4r+2\delta)}{(4r+\delta)\delta}-\frac{2}{(2r+\delta)}-\frac{4r^2}{(2r+\delta)^3}-\frac{2r^2(4r+2\delta)}{(4r+\delta)^2\delta^2}\right]\mathbf{n}
    \label{eq:vdw-force-particles}
\end{equation}
Since in the limit as $\delta$ approaches $0$ the magnitude of force (and potential energy) becomes infinite, a saturation distance $\delta_0$\nomenclature[M,d0]{\(\delta_0\)}{Hamaker saturation distance} is introduced. The magnitude of $\delta_0$ varies between 0.4 and 1 nm \cite{ranade1987adhesion}. Equation \ref{eq:vdw-force-particles} is used to simulate van der Waals attraction between two primary particles in the aggregate.

The attractive potential between a sphere and an infinite plane can be found by taking the limit of Equation \ref{eq:hamaker} as one of the particle radii approaches infinity:
\begin{equation}
    U=-\frac{A}{6}\left[\frac{r}{\delta}+\frac{r}{2r+\delta}+\ln\frac{\delta}{2r+\delta}\right]
    \label{eq:plane-potential}
\end{equation}
The expression for attractive force acting on a particle in the vicinity of the plane
is then obtained by differentiation of Equation \ref{eq:plane-potential}:
\begin{equation}
    \mathbf{f}=\frac{2A}{3}\left[\frac{r^3}{\delta^2\left(\delta+2r\right)^2}\right]\mathbf{n}
    \label{eq:vdw-force-particle-plane}
\end{equation}
Equation \ref{eq:vdw-force-particle-plane} is used to model attraction between a primary particle in the aggregate and either the substrate or the AFM tip in force spectroscopy simulations. We use $A=1.0\times10^{-19}\ \rm J$ for particle-particle and particle-substrate attraction, which is within the range of Hamaker constant values for graphite reported in literature \cite{krajina2014direct,tolias2018lifshitz}. We use $A=1.0\times10^{-18}\ \rm J$ for particle-tip attraction to ensure rapid adhesion of the aggregate to the tip during indentation.

\subsubsection{Capillary force representation}

\label{sec:capillary-force}

It is well-known that capillary forces exerted by liquid menisci on the pore walls inside porous materials cause adsorption-induced deformation \cite{gor2017adsorption,gor2024perspective}. The same capillary force drives soot restructuring, except the menisci form between primary particles in a fractal-like aggregate and not between the walls of concave pores. To simulate the capillary force that occurs in pairs of primary particles, we chose a sigmoid force-distance function:
\begin{equation}
    \mathbf{f}=\frac{1}{2}\psi\left[1-\tanh\left(\kappa\left(\lVert\mathbf{x}_j-\mathbf{x}_i\rVert-d_\mathrm{cutoff}\right)\right)\right]
    \label{eq:capillary-force-model}
\end{equation}
where $\psi$\nomenclature[M,p]{\(\psi\)}{maximum capillary force magnitude} is the maximum force magnitude (related to surface tension of the coating), $d_\mathrm{cutoff}$\nomenclature[M,dcutoff]{\(d_\mathrm{cutoff}\)}{capillary force cutoff distance} is inter-particle distance at which the force drops to zero (related to the amount of coating present on the aggregate), and $\kappa$\nomenclature[M,k]{\(\kappa\)}{sharpness of capillary force drop} is the sharpness of the drop that occurs at $d_\mathrm{cutoff}$. As $\kappa$ becomes large, the expression in Equation \ref{eq:capillary-force-model} becomes a step function. The reason why we chose a sigmoid dependence to simulate the capillary force is that it is a good mathematical model for a short-range force. For example, if $d_\mathrm{cutoff}=4r$, then only monomers that are separated by one monomer can interact. We use $\psi=1.0\times10^{-9}\ \rm N$ in this article, so that the magnitude of capillary force is of the same order as reported in literature \cite{farshchi2006adhesion}. In a more rigorous implementation, capillary force can be calculated in the model from the shape of a meniscus formed between two particles (\textit{i.e.} catenoid or globoid  \cite{ivanova2020kinetic}).

\subsubsection{Generation of aggregates}

\label{sec:fractals}

Aggregation naturally produces self-similar shapes that can be conveniently described through by the fractal scaling law \cite{jullien1987aggregation}:
\begin{equation}
    N=k_{\rm f}\left(\frac{R_{\rm g}}{r}\right)^{D_{\rm f}}
    \label{eq:fractal-law}
\end{equation}
where $N$ is the number of primary particles in the aggregate, $k_{\rm f}$\nomenclature[M,kf]{\(k_{\rm f}\)}{fractal pre-exponential factor} is the pre-exponential factor, $R_{\rm g}$\nomenclature[M,Rg]{\(R_{\rm g}\)}{aggregate radius of gyration} is the root mean square distance of primary particles from the center of mass of the aggregate, and $D_{\rm f}$\nomenclature[M,Df]{\(D_{\rm f}\)}{fractal dimension} is the fractal dimension. The two fractal parameters, $k_{\rm f}$ and $D_{\rm f}$, are determined experimentally from imaging \cite{lapuerta2006method} and are typically characteristic for all aggregates in a sample, while $N$ and $R_{\rm g}$ vary between different aggregates in one sample. Using the notion of fractals then allows to numerically generate aggregates, geometries of which are representative of real aggregates observed experimentally. Fresh soot aggregates have a $D_{\rm f}$ of $1.8$. As the aggregate acquires a coating and restructures, its $D_{\rm f}$ increases and approaches the limit of $\sim 2.2$ for thickly-coated aggregates \cite{wang2017fractal}.

A typical algorithm for numerical generation of fractal aggregates involves sequentially combining primary particles into clusters and subsequently combining sub-clusters into larger clusters in a way that the fractal scaling law (Equation \ref{eq:fractal-law}) holds until the desired aggregate size, $N$, is attained \cite{filippov2000fractal}. By using this procedure to generate aggregates with $D_{\rm f}$ of $1.8$, we get structures that are representative of real soot aggregates. In this study, we used the cluster-cluster aggregation (CCA) algorithm described by Filippov \textit{et al.} and implemented by Skorupski \textit{et al.} \cite{skorupski2014fast} as part of FLAGE software.

\section{Results and discussion}

The contact model described in this article was implemented in an in-house code. Technical details of the implementation are described in Supporting Information. Each elementary interaction model (non-bonded contact, bonded contact, and van der Waals attraction) were validated for compliance with conservation of energy and momentum. Validation was carried out by setting up simple simulations involving two or three particles with only the interaction models of interest enabled. Conserved quantities were recorded throughout the simulations. Descriptions of validation simulations and plots of conserved quantities are available in Supporting Information.

To put the developed framework to a test, two types of simulations are set up with the goal of qualitatively reproducing experimental results. The first simulation involves only the mechanical properties of aggregates and corresponds to the AFM indentation experiment conducted by Rong \textit{et al.} \cite{rong2004complementary} and also in our lab, where soot samples were collected on a silicon substrate and AFM indentation measurements were performed to determine the mechanical properties of those aggregates. The second simulation is a simplistic representation of restructuring of a thinly-coated soot aggregate, such as in experiments by Chen \textit{et al.} \cite{chen2016unexpected,chen2018single} (Figure \ref{fig:restructuring-illustration}).

\begin{figure}[p]
    \centering
    \includegraphics[width=0.95\textwidth]{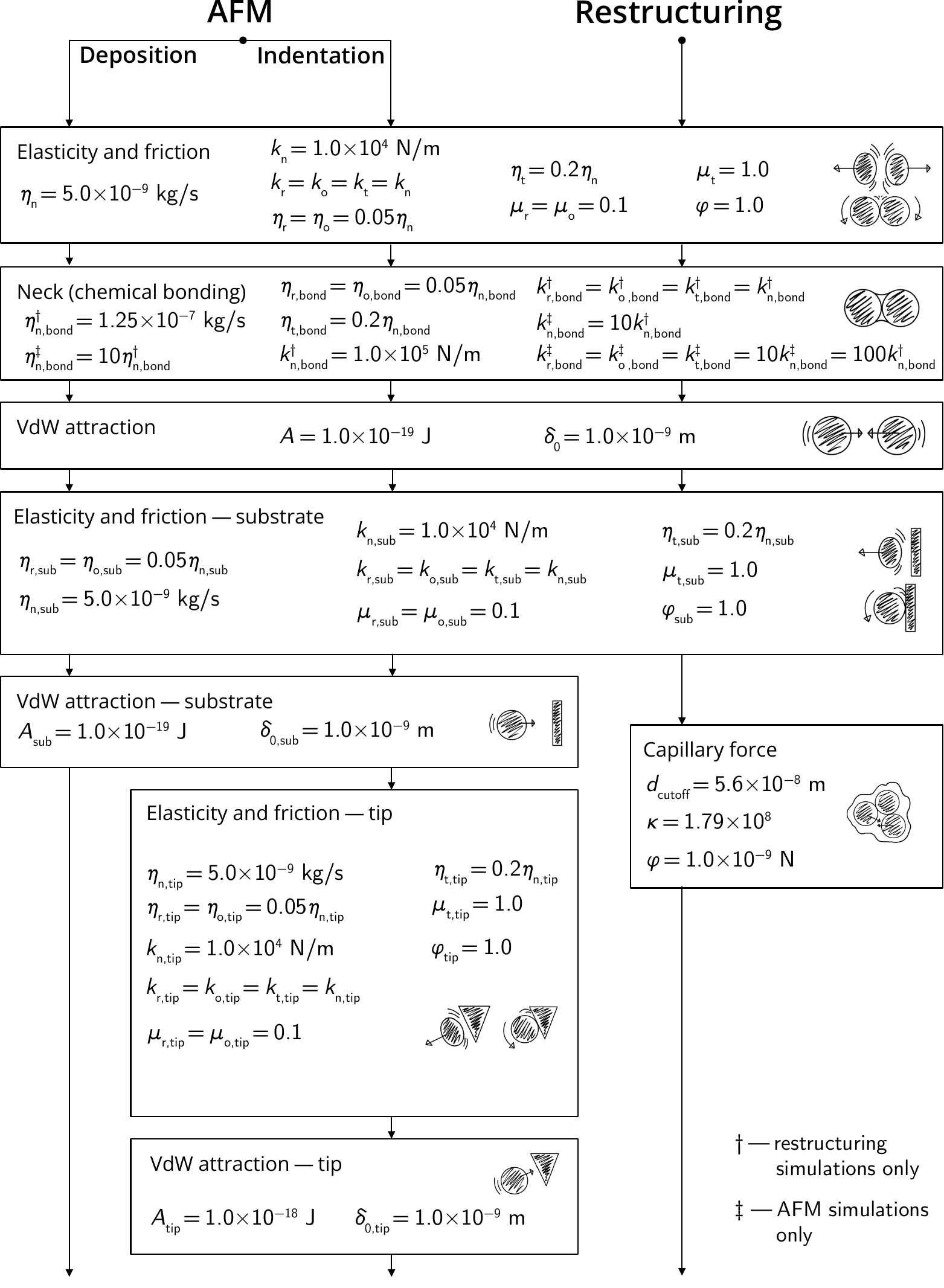}
    \caption{Force model parameters used in each simulation type}
    \label{fig:force-parameters}
\end{figure}

AFM deposition/indentation simulations were performed with an integration time step of $5.0\times 10^{-14}\ \rm s$ for a total simulation duration of $5.0\times 10^{-7}\ \rm s$. Restructuring simulations were performed with an integration time step of $5.0\times 10^{-15}\ \rm s$ for a total simulation duration of $5.0\times 10^{-8}\ \rm s$. Both AFM deposition/indentation and restructuring simulations were performed with a neighbor list update period of $20$ time steps, particle radius of $14\ \rm nm$, and particle material density of $1700\ \rm kg/m^3$. Verlet radius used to build neighbor lists was $4.2\times 10^{-8}\ \rm m$ for AFM indentation simulations and $7.0\times 10^{-8}\ \rm m$ for restructuring simulations. Parameters of the force models used in each simulation type are listed in the diagram in Figure \ref{fig:force-parameters}.

\subsection{AFM indentation simulation}

The AFM indentation simulation consists of two stages: deposition of an aggregate onto a substrate and indentation of a deposited aggregate with a solid tip.

\subsubsection{Aggregate deposition}

\begin{figure}[htp]
    \centering
    \includegraphics[width=\textwidth]{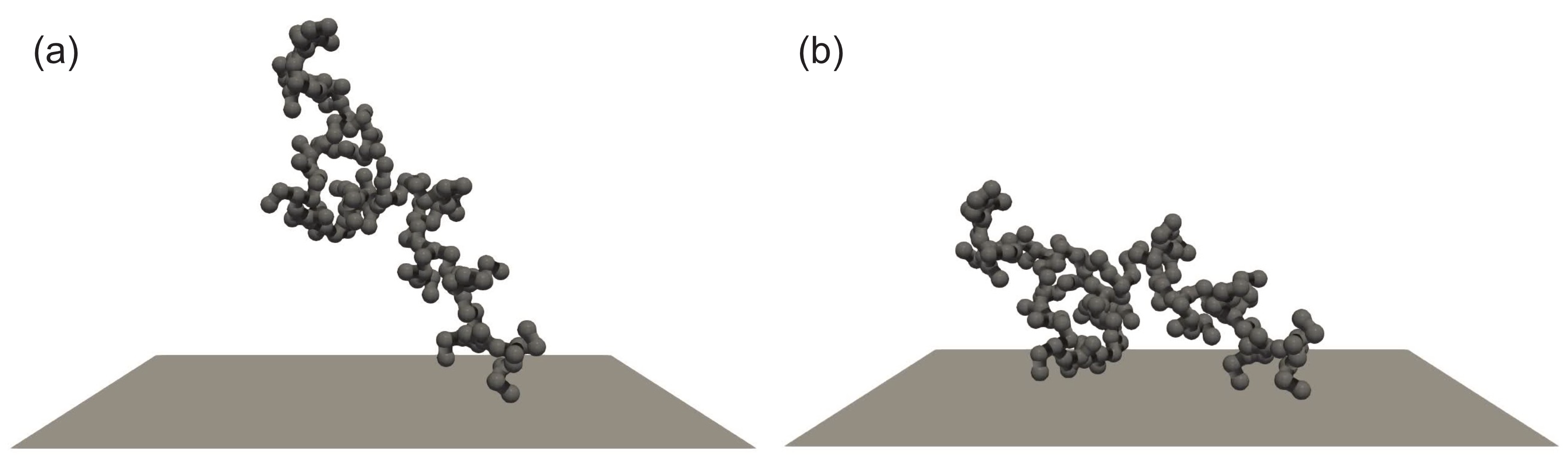}
    \caption{Initial and final states of an aggregate deposition simulation. (a) The aggregate is loaded, assigned an initial downward velocity, and positioned above the substrate. (b) The aggregate then collides with the substrate and adheres to it. A video file with a visualization of a deposition simulation is available in Supporting Information}
    \label{fig:deposition-illustration}
\end{figure}

This simulation stage begins with an aggregate, where all neighboring primary particles are bonded with necks. The substrate is introduced into the simulation as a static plane. It can exert forces on the granular system, like van der Waals attraction, normal force, and friction, but cannot accelerate itself. AFM tip is not present at this simulation stage. The substrate plane is normal to the z-axis and is located at $z=0$. A numerically generated fractal aggregate is loaded and positioned such that the lowest monomer (with respect to the z-axis) is only slightly above the substrate plane (Figure \ref{fig:deposition-illustration}a). Each primary particle is initialized with $1\ \rm m/s$ velocity in the negative-z direction. This is a reasonable deposition velocity at which no change in aggregate morphology upon impact is expected. e.g., as reported by Rothenbacher \textit{et al.} \cite{rothenbacher2008fragmentation}, fully necked aggregates do not fragment at impact velocities up to $300\rm \ m/s$.

In the course of the simulation, the aggregate approaches the substrate plane and collides with it. Adhesion occurs due to van der Waals attraction between the substrate plane and the primary particles in the aggregate and viscous dissipation of kinetic energy at the contact points during the collision of the aggregate with the substrate. The aggregate may continue its motion towards the substrate and rotate by inertia after one of the branches has adhered to the substrate. The motion continues until adhesion has occurred at multiple contact points (Figure \ref{fig:deposition-illustration}b). Then, the system becomes strongly anchored to the substrate.

\subsubsection{Aggregate indentation}

This simulation stage consists of an aggregate that is deposited onto the substrate. A preset fraction of neighboring primary particles are left with necked bonds, and necked bonds are replaced with non-bonded VdW contacts randomly in the remaining pairs. The substrate has the same representation and position as in the deposition simulations. AFM tip is introduced as a tetrahedron, consisting of solid triangle facets. Each triangle facet can exert forces on the granular system, like van der Waals attraction, normal force, and friction, but the facets themselves cannot accelerate. The triangle facets move along a pre-determined trajectory.

\begin{figure}[htp]
    \centering
    \includegraphics[width=\textwidth]{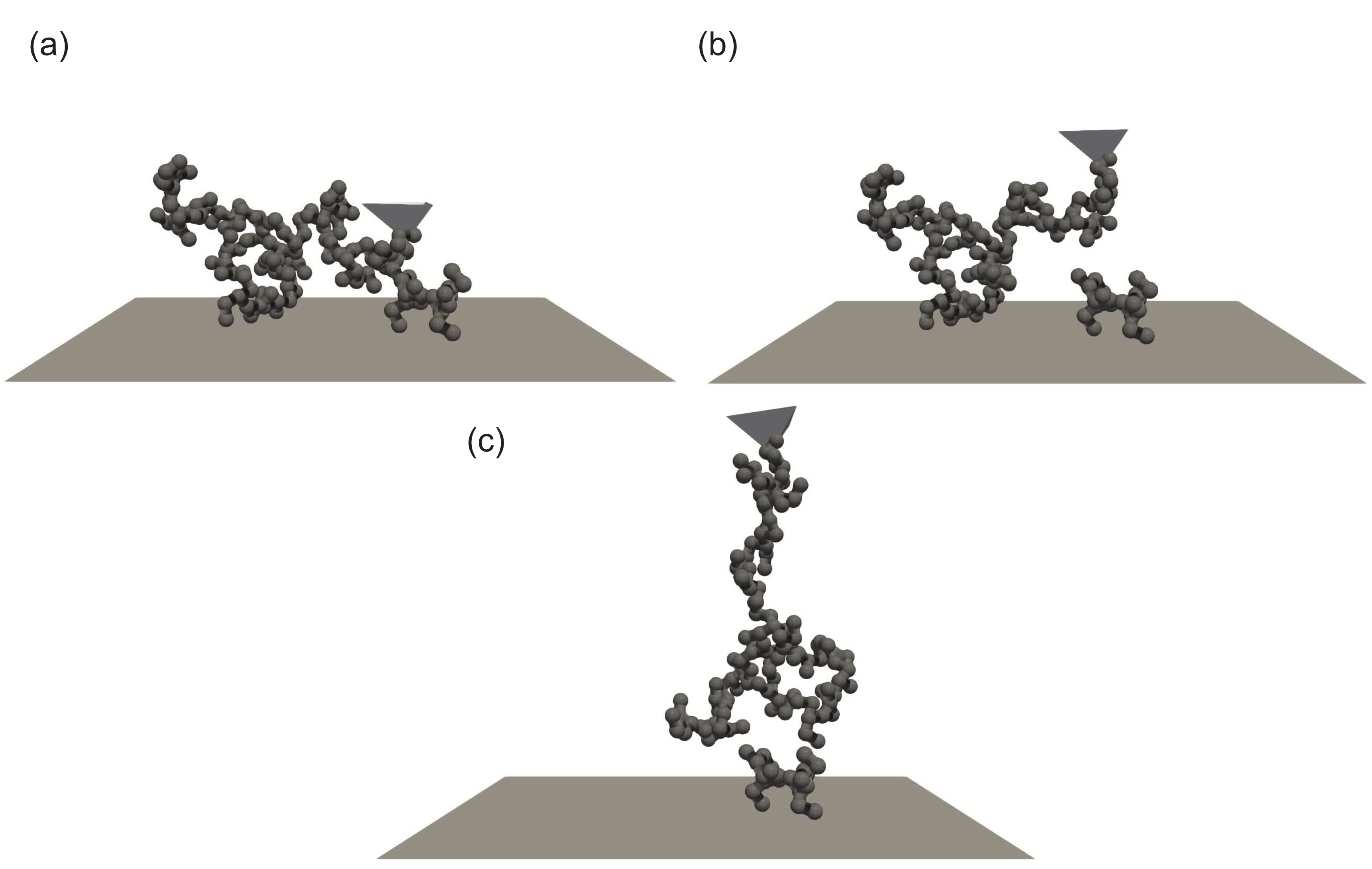}
    \caption{Initial, intermediate, and final states of an AFM indentation simulation. The AFM tip is positioned above a deposited aggregate, (a) is driven into the aggregate, and (b) retracted away from the aggregate. (c) Tension causes a part of the aggregate to separate from the system. A video file with a visualization of an indentation simulation is available in Supporting Information}
    \label{fig:indentation-illustration}
\end{figure}

The aggregate from the final data dump of a deposition simulation is loaded and the necessary fraction of bonds is established. A primary particle that is not hindered (along the z-axis) is randomly selected as the indentation point. The AFM tip is then positioned with its lowest vertex $0.5r$ above the selected primary particle and begins moving towards the primary particle (Figure \ref{fig:indentation-illustration}a). Soon after the tip has collided with the primary particle, it reverses its direction of motion and starts retraction. The motion of the AFM tip along the z-axis in indentation simulations is prescribed by following law (in $\rm m/s$):
\begin{equation}
    \dot{z}_\mathrm{tip}=1.5\tanh\left[{10^8(t-1.08\times 10^{-8})}\right]
\end{equation}
The velocity profile has a sigmoid shape. The tip moves with a constant velocity of $1.5\ \rm m/s$ towards the substrate, then at time $t=10^{-8}\ \rm s$\nomenclature[M,t]{\(t\)}{time} it smoothly reverses its motion and starts moving away from the substrate with a constant velocity of $1.5\ \rm m/s$. Due to strong adhesion, a branch of the aggregate attaches to and follows the tip (Figure \ref{fig:indentation-illustration}b). The force exerted by the AFM tip on the granular system is recorded as a function of tip height. It is important to note that in AFM force spectroscopy experiments, the probe moves at much lower velocities, on the order of microns per second, and the timescale of experiments is proportionally larger. However, running simulations for such long time periods is not feasible and numerical studies resort to stretching velocities on the order of meters per second \cite{laube2018new}.

In the course of the simulation, a part of the aggregate separates from the rest of the system along the broken necked bonds due to tension from the AFM tip. The separated part collides with and engages in multiple frictional contacts with the remaining aggregate pieces, until the piece adhered to the AFM tip finally separates from the rest of the system (Figure \ref{fig:indentation-illustration}c). These repeated frictional contacts result in the characteristic ``sawtooth'' profile, as reported by Rong \textit{et al.} \cite{rong2004complementary}.

\begin{figure}[htp]
    \centering
    \includegraphics[width=0.6\textwidth]{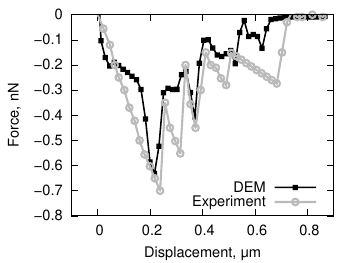}
    \caption{A simulated force-displacement curve (with the transfer function filter applied) plotted next to an example experimental force-displacement curve}
    \label{fig:indentation-plots}
\end{figure}

A simulated sawtooth curve is shown in Figure \ref{fig:indentation-plots}a. It was obtained from an indentation simulation with the transfer function described in Section \ref{sec:transfer-function} applied to smooth the recorded force. Figure \ref{fig:indentation-plots}b shows the simulated force-displacement curve next to an example of our experimental curve. The force spectroscopy experiment was performed similar to  Rong \textit{et al.} \cite{rong2004complementary}, with a Bruker Dimension Icon AFM under ambient conditions using a cantilever with the spring constant of $0.4\ \rm N/m$ and retraction velocity of $1\ \rm \mu m/s$. Soot aggregates were electrostatically size-selected to $240\ \rm nm$, which corresponds to about 120 primary particles per aggregate. A manuscript describing the experimental AFM measurements is in preparation.

The simulated force-displacement curve has the expected sawtooth profile. The magnitude of force peaks, their number, and the displacement at which the aggregate is completely severed are comparable between experimental and simulated data. The outcome of indentation simulations is mostly affected by (1) initial morphology of the aggregate and (2) the Hamaker constant used to model van der Waals attraction between monomers. Aggregate morphology controls the number of peaks seen on a force-displacement curve, as the probability of multiple collisions between branches occurring is higher in a larger and more lacey aggregate. Hamaker constant controls the magnitude of the force peaks on the force-displacement curve. The good agreement between simulations and experiments can be explained by us using realistic fractal parameters (Section \ref{sec:fractals}), a Hamaker constant value for graphite consistent with literature (Section \ref{sec:non-contact-forces}), and aggregates with approximately the same number of primary particles as in our experiments. Varying the number of monomers in the aggregate may affect the total number of sawtooth peaks on the force-displacement curve.


\subsection{Restructuring simulation}

\begin{figure}[htp]
    \centering
    \includegraphics[width=\textwidth]{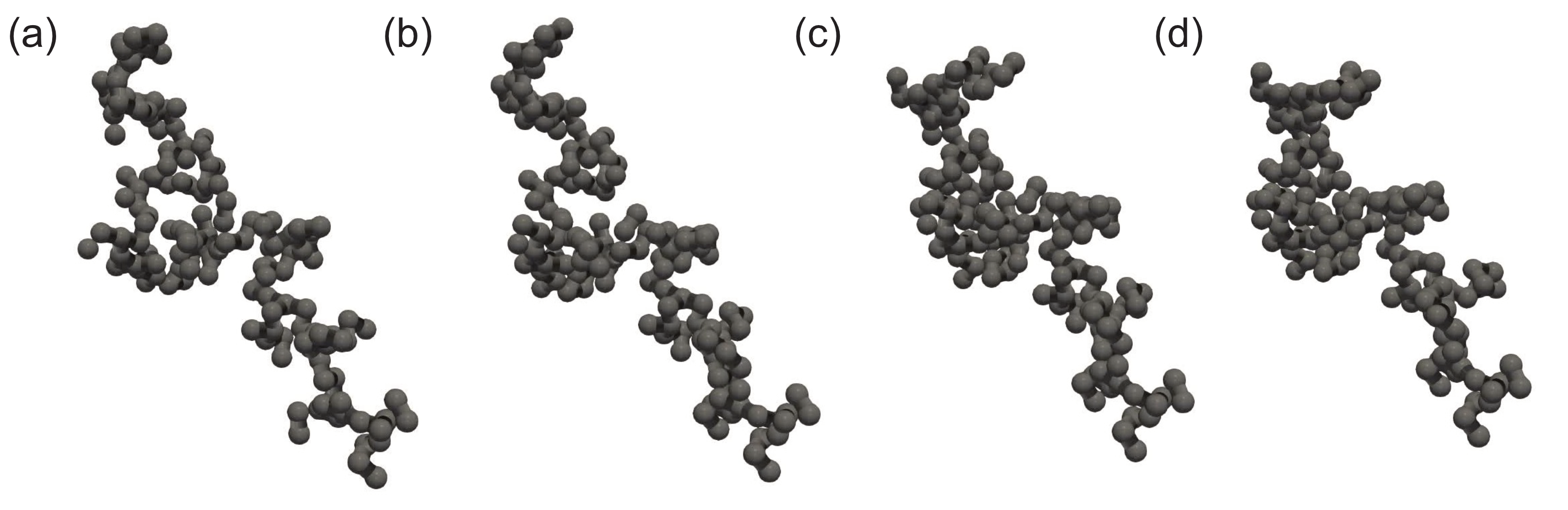}
    \caption{Initial through final states of an aggregate restructuring simulation. (a-d) Capillary forces cause short, peripheral branches to collapse first. Backbone branches rearrange at a slower rate and the aggregate reaches a stable configuration eventually, where all the non-bonded contacts are locked. A video file with a visualization of a restructuring simulation is available in Supporting Information}
    \label{fig:restructuring-illustration}
\end{figure}

The simulation consists of an aggregate with necked connections between primary particles and a capillary force acting on it. A preset fraction of neighboring primary particles are left in non-bonded contacts with each other to allow for restructuring. In the course of the simulation, the aggregate undergoes compaction. The shortest branches collapse first, while longer branches take time to accelerate. The change in morphology of an aggregate during a restructuring simulation is presented in Figure \ref{fig:restructuring-illustration}.

\begin{figure}[htp]
    \centering
    \includegraphics[width=\textwidth]{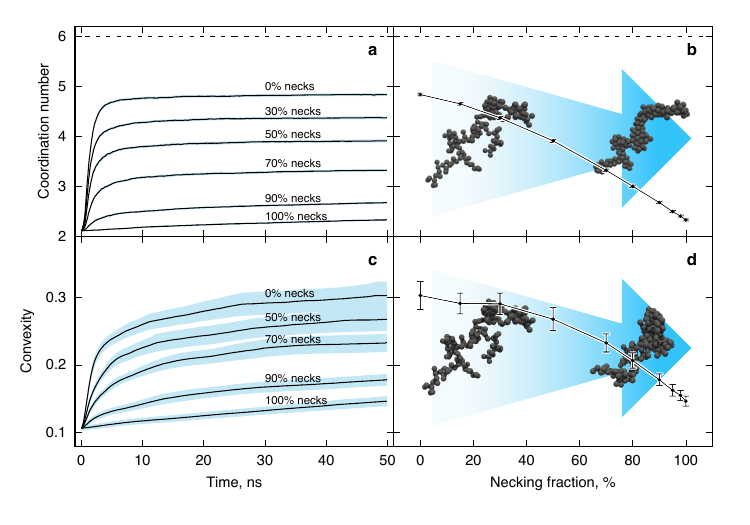}
    \caption{Evolution in morphology of soot aggregates undergoing restructuring due to capillary force. (a) Aggregate-average coordination number of 20 restructuring aggregates as a function of time, (b) average terminal coordination number as a function of aggregate necking fraction. The dashed line corresponds to the limiting coordination number of 6.0 for bulk random-jammed packing of frictionless spheres \cite{silbert2010jamming}. (c) Average convexity of 50 restructuring aggregates as a function of time, (d) average terminal convexity as a function of aggregate necking fraction. Inset in sub-plot b with an arrow and aggregates illustrates local compaction and inset in sub-plot d illustrates global compaction. Shaded regions and error bars are 95\% confidence interval}
    \label{fig:restructuring-plots}
\end{figure}

Average coordination number of primary particles in the aggregate was chosen as the metric to track the progress of restructuring. It is defined as the average number of neighbors that a primary particle has and is computed as,
\begin{equation}
    c=\frac{1}{N}\sum_{i=1}^NN_{\mathrm{nb},i}
\end{equation}
where $c$\nomenclature[M,c]{\(c\)}{aggregate coordniation number} is the average coordination number of the aggregate and $N_{\mathrm{nb},i}$\nomenclature[M,Nnb]{\(N_{\rm nb}\)}{number of neighbors of a particle} is the number of neighbors that particle $i$ has. For the purpose of finding $N_{\mathrm{nb},i}$, we consider as neighbors two particles, which are separated by less that $1\%$ of primary particle diameter. In restructuring simulations, 50 different aggregates with the same size (150 primary particles) and fractal parameters ($1.8$ fractal dimension and $1.4$ pre-exponential factor) were restructured with the same coating force for the same duration. Average coordination numbers of aggregates with various necking fractions are presented in Figure \ref{fig:restructuring-plots}a,b. Coordination number is descriptive of local morphological changes in an aggregate (inset in Figure \ref{fig:restructuring-plots}b). We also compute convexity of the restructuring aggregates, which in three dimensions is defined as the ratio of the volume of an aggregate to the volume of the convex hull constructed around the aggregate \cite{cgal:hs-ch3-24a}. Convexity is descriptive of global morphological changes in the aggregate (inset in Figure \ref{fig:restructuring-plots}d). Convexity data is presented in Figure \ref{fig:restructuring-plots}c,d. Three-dimensional convexity presented in Figure \ref{fig:restructuring-plots} cannot be trivially related to experimental convexity derived from images, as in experimental studies convexity is two-dimensional (ratio of aggregate projected area to the area of the convex polygon) and is an inherently different parameter. Terminal two-dimensional convexity can be estimated by averaging convexities of aggregate projections onto the $xy$, $xz$, and $yz$ planes. For aggregates with necking fractions of $0\%$ and $70\%$, terminal 2D convexity was found to be $0.81\pm0.03$ and $0.80\pm0.03$ respectively. Computed 2D convexities lie in the range from $0.75$ to $0.87$, reported for restructured aggregates in experimental studies \cite{chen2016unexpected,chen2018single}.

In all simulations, the coordination number is a good indicator of restructuring progress. As it reaches its maximum value, restructuring slows down and stops. Maximum coordination number correlates with the fraction of necks present in the aggregate. Fewer necks allow for more compaction and a higher coordination number is attained by the aggregate at the end of restructuring. A slight increase in coordination number (under $10\%$) is observed for aggregates with $100\%$ necking due to elastic deformation of necks. Aggregates with no necks attain the highest coordination number, $c=4.48$. This is well below the limiting coordination number of $6.0$ for bulk random-jammed frictionless packing, due to inter-particle friction being enabled and a large fraction of particles being on the surface of the aggregate \cite{silbert2010jamming}. Full compaction occurs within tens of nanoseconds. The timescale of restructuring depends mostly on the magnitude of the driving force, which was selected by us based on literature (Section \ref{sec:capillary-force}). Convexity takes longer to reach steady state and exhibits more variance between different aggregates. Coordination number reaches steady state sooner than convexity because it characterizes local compactness of the aggregate and local compaction occurs faster than global movement of long aggregate branches.

\subsection{Limitations and potential future developments}

One of the limitations of the capillary force model used in this study is that it does not take into account hysteresis related to formation / breakage of liquid bridges \cite{willett2003effects}. This capillary force model cannot be used to model thickly coated soot, where the liquid layer covers the entire aggregate. Particularly, in the latter case, the moving air-liquid interface of an evaporating coating droplet pushes all particles towards the center of the droplet instead of a causing particle-particle attraction, as described by the present model. Future work will involve development of a more elaborate representation of coating materials and parametrization of the restructuring model based on experiments. An algorithm for brittle fracture of necks will be implemented and neck strengths will be parametrized. With those additions implemented on top of the DEM contact model presented in this study, the restructuring model in conjunction with the capillary condensation model developed by our group will be used to study evolution of soot composition, morphology, and optical properties under atmospheric conditions. The DEM contact model will also be applied to simulating mechanical properties of industrial carbon blacks incorporated into various media.

\section{Conclusion}

We have developed a DEM contact model for mechanics of soot aggregates. Elementary interaction models were validated for consistency. The contact model was parametrized based on AFM spectroscopy experiments and literature data on carbon Hamaker constant. Restructuring simulations were conducted using a simplistic model for capillary force. The magnitude of capillary force used to drive restructuring in this study was on the order of nano-Newtons, which is comparable to values reported in literature for similar systems \cite{farshchi2006adhesion}. With this order of capillary force, the timescale of restructuring is tens of nanoseconds. Hence, the rate of restructuring is much faster than the rate of condensation in saturator-based experiments (milliseconds \cite{ivanova2020kinetic,chen2018single}), chamber-based experiments (minutes \cite{wittbom2014cloud,khalizov2013role}), and the atmosphere (hours \cite{riemer2004soot,zhang2015long,riemer2010estimating}). Therefore, the rate of soot compaction is controlled by the rate of coating formation by vapor condensation, not the rate of restructuring. This means that condensation and restructuring simulations can be decoupled, which significantly simplifies the computational burden for future condensation-driven restructuring modeling.

Simulations also show that the outcome of the restructuring process is controlled by the number of necks present in the aggregate. In an aggregate with near-full necking, where only a few necks are broken, global restructuring occurs due to movement of bonded branches as a whole. In an aggregate with most of the necks broken, local restructuring is prevalent, because exterior particles quickly collapse onto backbone branches, interlock, and inhibit further restructuring. The fact that major restructuring can occur with only one or two necks broken in the aggregate can explain the highly compact thinly-coated soot, where the amount of condensate is only sufficient to break a few necks, observed in field measurements \cite{bhandari2019extensive,china2015morphology,wang2024variability} and in laboratory studies \cite{bambha2013effects}. Likewise, minor restructuring of thickly coated soot, where the amount of condensate is sufficient to break most necks in the aggregate, observed in some studies \cite{cross2010soot,ghazi2013coating}, can be attributed to fast condensation followed by interlocking of primary particles.

\section*{Authors' contributions}

Egor V. Demidov: Software, Writing -- review \& editing, Writing -- original draft, Visualization, Validation, Methodology, Investigation, Formal analysis, Conceptualization. Gennady Y. Gor: Writing -- review \& editing, Supervision, Resources, Project administration, Methodology, Funding acquisition, Formal analysis, Conceptualization. Alexei F. Khalizov: Writing -- review \& editing, Supervision, Resources, Project administration, Methodology, Funding acquisition, Formal analysis, Conceptualization.

\section*{Data availability}

Supporting information provided with this manuscript consists of a PDF document with a description of the implementation details of the model and validation steps undertaken; and an archive with videos of select deposition, indentation, and restructuring simulations; an archive with input files that can be used to reproduce all data used in this article. Software is available in a \href{https://github.com/egor-demidov/soot-dem}{GitHub repository}. A tutorial on how to run restructuring simulations using our software is available on \href{https://www.edemidov.com/posts/soot-restructuring}{Egor Demidov's web page}.

\section*{Acknowledgment}

A.K., G.G., and E.D. acknowledge the U.S. National Science Foundation, grant AGS-2222104. We thank Ali Hasani for the experimental force-displacement curve used in Figure \ref{fig:indentation-plots} and for the TEM image of a soot aggregate used in Figure \ref{fig:contact-types}.

\appendix
\section*{Appendix}
\renewcommand{\thesection}{\Alph{section}}
\renewcommand{\thesubsection}{\Alph{section}.\arabic{subsection}}
\setcounter{section}{1}

\subsection{AFM transfer function}

\label{sec:transfer-function}

A filter is needed to reduce the high-frequency noise on simulated force-displacement curves. The real AFM cantilever is a damped oscillator system \cite{butt2005force} with nonzero mass. Due to the inertia of the cantilever, an infinitesimal force impulse applied to the cantilever will not cause it to deflect by a detectable distance and will not be measured by the instrument. Here, we derive a transfer function that relates the applied force to the force measured by the AFM instrument.

Let the AFM cantilever be a second-order system:
\begin{equation}
    m_{\rm c}\ddot{u}+\eta_{\rm c}\dot{u}+k_{\rm c}u=f(t)
\end{equation}
where $m_{\rm c}$\nomenclature[P,mc]{\(m_{\rm c}\)}{effective mass of an AFM cantilever} is the effective mass of the cantilever, $\eta_{\rm c}$\nomenclature[P,nc]{\(\eta_{\rm c}\)}{damping coefficient of an AFM cantilever} is the damping coefficient of the cantilever, $k_{\rm c}$\nomenclature[P,kc]{\(k_{\rm c}\)}{stiffness of an AFM cantilever} is the effective stiffness of the cantilever, $u$\nomenclature[P,u]{\(u\)}{displacement of an AFM cantilever} is the displacement of the cantilever from the equilibrium position, and $f(t)$ is the external force applied to the cantilever. The instrument determines force by measuring $u$ by detecting deflection of a laser beam that is being reflected from the cantilever. Then instrument software computes the apparent force using Hooke's law, $g=k_{\rm c}u$, and records it. We can establish a relationship between the true force applied to the AFM probe and the apparent force that is recorded by the instrument as:
\begin{equation}
    \frac{m_{\rm c}}{k_{\rm c}}\ddot{g}+\frac{\eta_{\rm c}}{k_{\rm c}}\dot{g}+g(t)=f(t)
    \label{eq:afm-ode}
\end{equation}
Assuming that the cantilever is critically damped,
\begin{equation}
    \eta_{\rm c}=2\sqrt{m_{\rm c}k_{\rm c}}
\end{equation}
and by taking a Laplace transform of Equation \ref{eq:afm-ode}, we can arrive at the expression for the transfer function of the instrument:
\begin{equation}
    \frac{G(s)}{F(s)}=\frac{\omega_0^2}{\left(\omega_0+s\right)^2}
    \label{eq:afm-transfer}
\end{equation}
where $\omega_0$ is the natural angular frequency of the undamped oscillator given by:
\begin{equation}
    \omega_0=\sqrt{\frac{k_{\rm c}}{m_{\rm c}}}
\end{equation}
By exploring the limits of the transfer function (Equation \ref{eq:afm-transfer}), we can see that for a large value of $\omega_0$, the input is preserved, and for a small value of $\omega_0$, the input is smoothed.
\begin{equation}
    \lim_{\omega_0\rightarrow\infty}\frac{\omega_0^2}{(s+\omega_0)^2}=1\quad\Rightarrow\quad\text{input is preserved for large $\omega_0$}
\end{equation}
\begin{equation}
    \lim_{\omega_0\rightarrow 0}\frac{\omega_0^2}{(s+\omega_0)^2}=0\quad\Rightarrow\quad\text{input is attenuated for small $\omega_0$}
\end{equation}


\printbibliography

\end{document}